\documentclass[print,epsfig,doublecol,floats,showpacs]{revtex4}
\usepackage{graphicx}
\usepackage{dcolumn}
\usepackage{bm}
\usepackage{amssymb}
\usepackage{amsmath}
\usepackage{epsfig}
\usepackage{graphicx}
\usepackage{dcolumn}
\begin{document}
\title{Multiparty-controlled remote preparation of four-qubit cluster-type entangled states}
\author{Dong Wang$^{a,b,c,}$\footnote{{dwang@ahu.edu.cn (D. Wang)}}, Liu Ye$^{b,}$\footnote{{yeliu@ahu.edu.cn (L. Ye)}},
Sabre Kais$^{a,d,}$\footnote{{kais@purdue.edu (S. Kais)}}}
\affiliation{${\ ^a}$ Department of Chemistry and Birck Nanotechnology Center,
Purdue University, West Lafayette, IN 47907, USA\\
${\ ^b}$ School of Physics \& Material Science, Anhui University, Hefei
230601, China\\
${\ ^c}$ National Laboratory for Infrared Physics, Shanghai Institute of Technical Physics, Chinese Academy of Sciences, Shanghai 200083, China
\\
${\ ^d}$ Qatar Environment and Energy Research Institute, Qatar Foundation, Doha, Qatar}

\begin{abstract}
We present a strategy for implementing multiparty-controlled remote state preparation (MCRSP)
for a family of four-qubit cluster-type states with genuine entanglements while employing, Greenberg-Horne-Zeilinger-class states as quantum channels. In this scenario, the encoded information is transmitted from the sender to a spatially separated  receiver via the control of multi-party. Predicated on the collaboration of
all participants, the desired state can be entirely restored within the receiver's place with high success
probability by application of appropriate local operations
and necessary classical communication . Moreover, this proposal for MCRSP can be faithfully achieved with unit total success probability when the quantum channels are distilled to maximally entangled ones.

\end{abstract}

\maketitle

\section{Introduction}
An important focus in the field of quantum information processing (QIP) has been the secure and faithful
transmission of information from one node of quantum network to another non-local node with finite
classical and quantum resources. Quantum teleportation (QT)
originated from the  pioneering work of Bennett \cite{Bennett} is one application of non-local physics which may
accomplish such a task. the central idea of QT is to deliver magically quantum information
without physically transporting any particles from the sender to the receiver by means of an established entanglement.
Apart from QT there exists another such efficient method, the so-called remote state preparation
(RSP)~\cite{H.K., A.K.Pati, C.H.}. RSP allows for the transfer of arbitrary known quantum states from a sender (Alice)
to a spatially distant receiver (Bob), provided that the two parties share an entangled state
and may communicate classically. Although both QT and RSP are able to achieve the task of
information transfer~\cite{ZhuJing,Kais,KaisSabre}, there are some subtle differences between QT and RSP which
can be summarized as follow: (i) Precondition.
In RSP, the sender of the states is required to be completely knowledge about the prepared state. In contrast,
neither the sender nor the receiver necessarily
possesses any knowledge of the information associated with the
teleported states in QT. (ii) State existence. The state to be teleported initially
inhabits a physical particle in the context of QT, while this is not required in RSP. That is to say,
the sender in RSP is full aware of the information regarding the desired state, without any particle in such a state within his possession.
(iii) Resource trade-off. Bennett \cite{C.H.} has shown that quantum and classical resources can
be traded off in RSP but cannot in QT. In standard teleportation, an unknown quantum state is sent via a
quantum channel, involving 1 ebit, and 2 cbits for
communication. In contrast, if the teleported state is known to the
sender prior to teleportation, the required resources can be reduced to 1 ebit and
1 cbit in RSP at the expense of  lower success probability, half of that in QT. However,
Pati \cite{A.K.Pati} has argued that for special ensemble
states (e.g., states on either the equator or great polar circle of the Bloch sphere)
RSP requires less classical information than teleportation with the same unitary success probability.

Owing to its importance in QIP, RSP has received great attention and a large number of theoretical investigations
 have been proposed \cite{Devetak,Leung1,Berry,Kurucz1,Hayashi,Abeyesinghe,Ye,Yu,Huang,Xia1,
 Nguyen1,Luo1,Xiao-QiXiao22,Qing-QinChen4,Nguyen2,Luo2,PingZhou6,YanXia3,You-BanZhan6,
MingJiang2,Xiu-BoChen2,Zhi-HuaZhang,Dong1,Liu1,Liu2,Wang1,
 Dai1,Dai2,Mikami,Paris,Kurucz2}.
Specifically, there have been investigations concerning: low-entanglement RSP \cite{Devetak},
optimal RSP \cite{Leung1}, oblivious RSP \cite{Berry,Kurucz1}, RSP without oblivious conditions \cite{Hayashi},
generalized RSP \cite{Abeyesinghe}, faithful RSP \cite{Ye},
 joint RSP (JRSP) \cite{Xia1,Nguyen1,Luo1,Xiao-QiXiao22,Qing-QinChen4,Nguyen2,Luo2,PingZhou6,
 YanXia3,You-BanZhan6,MingJiang2,
 Xiu-BoChen2,Zhi-HuaZhang}, Multi-controlled joint RSP \cite{Dong1}, RSP for many-particle states
\cite{Yu,Huang,Liu1,Liu2,Wang1,Dai1,Dai2}, RSP for qutrit states \cite{Mikami}
and continuous variable RSP in phase space \cite{Paris,Kurucz2}. While, several RSP proposals by means of different physical systems
have been experimentally demonstrated
as well \cite{Peng,Xiang,Peters,Jeffrey,Liu4,Wu1,JulioT.Barreiro}. For examples,
Peng {\it et al.} investigated a RSP scheme using NMR \cite{Peng},
Xiang {\it et al.} \cite{Xiang} and Peters {\it et al.} \cite{Peters} proposed other two RSP
schemes using spontaneous parametric down-conversion. Julio {\it et al.} \cite{JulioT.Barreiro}
reported the remote preparation of two-qubit hybrid entangled
states, including a family of vector-polarization beams; where single-photon states are encoded in the photon
spin and orbital angular momentum, and then the desired state is reconstructed
by means of spin-orbit state tomography and
transverse polarization tomography.

Recently, many authors proceed to focus on RSP for cluster-type state by exploring various novel methods
\cite{Y.B,D.Wan6,D.Wang5,K.Hou333,Y.B.11,K.Hou}; because
cluster states are one of the most important resources in quantum information processing
and can be efficiently applied to
information processing tasks, such as: quantum teleportation \cite{P.P.}, quantum dense
coding \cite{X.W,C.W.Tsai}, quantum secret sharing \cite{H.K.Lau}, quantum computation \cite{Y.Wang}, and quantum
correction \cite{J.Joo}. In general, a cluster-state is expressed as
\begin{equation}
\begin{split}
|{\Omega}_N\rangle= \frac1{2^{N/2}}\bigotimes_{s=1}^{N}(|0\rangle_sZ_{(s+1)}+|1\rangle_s),
\end{split}
\end{equation}
with the conventional use of $Z$ is a pauli operator and $Z_{N+1}\equiv1$. It has been shown that one-dimensional
$N$-qubit cluster states are generated in arrays of $N$
qubits mediated with an Ising-type interaction. It may easily be seen that the state
will be reduced into a Bell state for $N=2$ (or 3); the
cluster states are equivalent to Bell states (or Greenberger-Horne-Zeilinger (GHZ) states) respectively under stochastic
local operation and classical communication (LOCC). Yet
when $N>3$, the cluster state and the $N$-qubit GHZ state
cannot be converted to each other by LOCC. When $N=4$, the four-qubit cluster-state
is given by
\begin{equation}
\begin{split}
|{\Omega}_4\rangle =\frac{1}{2}(|0000\rangle+|0011\rangle+|1100\rangle-|1111\rangle).
\end{split}
\end{equation}

In this work our aim is to examine the implementations of multiparty-controlled remote state preparation (MCRSP)
for a family of four-qubit cluster-type entangled states with the aid of general quantum channels \cite{Y.B,D.Wan6,D.Wang5,K.Hou333,Y.B.11,K.Hou}.

The paper is structured as follows: in the next section, we present the  MCRSP scheme for
four-qubit cluster-type entangled states with multi-agent control by the utilization of
GHZ-class entanglements as quantum channels. The results show that the desired state can be faithfully
reconstructed within Bob's laboratory with high success probability.
Moreover, the required classical communication cost (CCC) and total success probability
(TSP) will be discussed. Finally, features of our proposed scheme are
detailed followed by a conclusion section.

\section{MCRSP for four-qubit cluster-type entangled states}
Suppose there are $(m+n+2)$ authorized participants, say, Alice, Bob, Charlie$_1$, $\cdots$, Charlie$_n$, Dick$_1$, $\cdots$, and Dick$_m$ $({\rm where} \ m,n\geq1)$.
To be explicit, Alice is the sender of the desired state, Bob is the receiver, and Charlie$_i$ and
Dick$_j$ are truthful agents. Now, Alice would like to assist Bob remotely in the preparation of a four-qubit cluster-type entangled state
\begin{equation}
\begin{split}
|{{P}}\rangle=\alpha|0000\rangle+\beta{e^{i\varphi_0}}|0011\rangle+\gamma{e^{i\varphi
_1}}|1100\rangle+\delta{e^{i\varphi_2}}|1111\rangle,
\end{split}
\end{equation}
with the control of the agents, where $\alpha$, $\beta$, $\gamma$, $\delta$ and $\varphi_i$ are real-valued, satisfy the normalized condition $\alpha^2+\beta^2+\gamma^2+\delta^2=1$, and $\varphi_i\in[0,2\pi]$.
In order to obtain MCRSP, Alice, Bob, Charlie$_i$ and Dick$_j$  share previously generated genuine quantum resources -- i.e., GHZ entanglements -- which are given by
\begin{equation}
\begin{split}
|\Upsilon^{(1)}\rangle_{A_1A_2B_1B_2C_{1}\cdots{C_{n}}}=\sum_{k}^{0,1}a_k|k\rangle^{\otimes(n+4)}_{A_1A_2B_1B_2C_{1}\cdots{C_{n}}},
\end{split}
\end{equation}
and
\begin{equation}
\begin{split}
|\Upsilon^{(2)}\rangle_{A_3A_4B_3B_4D_{1}\cdots{D_{m}}}=\sum_{l}^{0,1}b_l|l\rangle^{\otimes(m+4)}_{A_3A_4B_3B_4D_{1}\cdots{D_{m}}},
\end{split}
\end{equation}
respectively, without loss of generality ${a}_1,{b}_1\in\mathbb{R}$, and these bounds $|a_0|\geq|{a}_1|$ and $|b_0|\geq|{b}_1|$
are maintained. Initially, qubits $A_1$, $A_2$, $A_3$ and $A_4$ are sent to Alice,
qubits $B_1$, $B_2$, $B_3$ and $B_4$ to Bob, $C_{i}$ to Charlie$_i$ $(i\in\{1,\cdots,n\})$ and $D_{j}$ to Dick$_j$ $(j\in\{1,\cdots,m\})$.

For implementing MCRSP, the procedure can be divided into the following steps:

${\bf Step\ 1.}$
Firstly, Alice makes a two-qubit projective measurement on her qubit pair ($A_1,A_3$)
under a set of complete orthogonal basis vectors $\{|{\mathcal{L}}_{ij}\rangle\}$ composed of computational basis $\{|00\rangle,|01\rangle,|10\rangle,|11\rangle\}$, which can be written as
\begin{equation}
 \begin{array}{c}
        \left(|{{\mathcal{L}}}_{00}\rangle, |{{\mathcal{L}}}_{01}\rangle ,|{{\mathcal{L}}}_{10}\rangle ,|{{\mathcal{L}}}_{11}\rangle \right)^T
          ={\cal Q} \left(|00\rangle, |01\rangle ,|10\rangle ,|11\rangle \right)^T,
            \end{array}
 \end{equation}
  where,
\begin{equation}
{\cal Q}=\left(\begin{array}{cccc}
\alpha & \beta{e^{-i\varphi_0}} & \gamma{e^{-i\varphi_1}} & \delta{e^{-i\varphi_2}} \\
\beta & -\alpha{e^{-i\varphi_0}} & \delta{e^{-i\varphi_1}} & -\gamma{e^{-i\varphi_2}} \\
\gamma & -\delta{e^{-i\varphi_0}} & -\alpha{e^{-i\varphi_1}} & \beta{e^{-i\varphi_2}} \\
\delta & \gamma{e^{-i\varphi_0}} & -\beta{e^{-i\varphi_1}} & -\alpha{e^{-i\varphi_2}} \\ \end{array}\right).
\end{equation}

Since the total systemic state taken as quantum channels can be described as
\begin{equation}\begin{split}
&|\Psi_{T}\rangle=|{\Upsilon}^{(1)}\rangle_{A_1A_2B_1B_2C_{1}\cdots{C_{n}}}\otimes|{\Upsilon}^{(2)}\rangle_{A_3A_4B_3B_4D_{1}\cdots{D_{m}}}  \\
&=\sum_{i,j}^{0,1}|{\cal L}_{ij}\rangle_{A_1A_3}\otimes|{\cal X}_{ij}\rangle_{A_2A_4B_1B_2B_3B_4C_1\cdots{C_n}D_1\cdots{D_m}},
\end{split}\end{equation}
where the non-normalized state $|{\cal X}_{ij}\rangle\equiv{_{A_1A_3}}\langle{\cal L}_{ij}|\Psi_T\rangle\
(i,j=0,1)$ is obtained with probability of $1/{{\cal N}_{ij}^2}$, where ${\cal N}_{ij}$ corresponds to the normalized
parameter of state $|{\cal X}_{ij}\rangle$.
\begin{figure*}
\centering
\includegraphics[width=15cm,height=6.5cm]{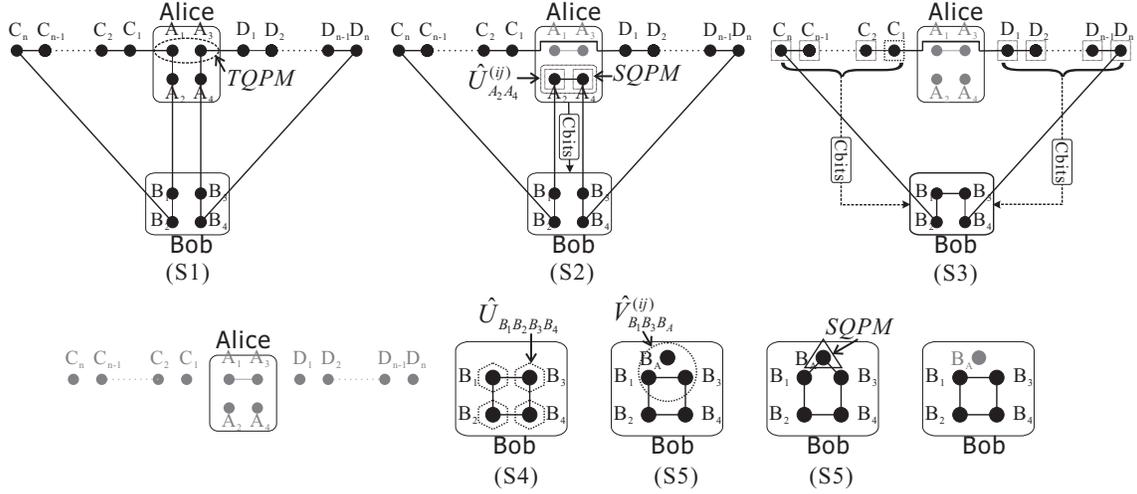}
\caption{Schematic diagram for  MCRSP implementation. The procedure is explicitly decomposed as above Figures (S1)$\sim$(S5).
The ellipse represents two-qubit projective measurement (TQPM) under the set of basis vectors $\{|{\cal L}_{ij}\rangle\}$;
the square represents single-qubit projective measurement (SQPM) under the set of basis vectors $\{|\pm\rangle\}$; rectangle represents
operating a bipartite collective unitary transformation ${\hat{U}}_{A_2A_4}^{(ij)}$;the triangle
represents SQPM under the set of basis vectors $\{|0\rangle,|1\rangle\}$; the sexangle represents performing single-qubit unitary transformations
$\hat{U}_{B_1B_2B_3B_4}$ on Bob's qubits;
the circle represents making a triplet collective unitary operation $\hat{V}_{B_1B_3B_A}^{(ij)}$; Cbits represents classical information communication.}
\label{fig.1}
\end{figure*}

${\bf Step\ 2.}$
According to her own measurement outcome $|{\cal L}_{ij}\rangle$, Alice makes an appropriate unitary operation
$\hat{U}_{A_2A_4}^{(ij)}$ on her remaining qubit pair ($A_2$, $A_4$) under the ordering basis
$\{|00\rangle,|01\rangle,|10\rangle,|11\rangle\}$, which is accordingly one of
\begin{equation}
\hat{U}_{A_2A_4}^{(00)}={\rm diag}(1,1,1,1),
\end{equation}
\begin{equation}
\hat{U}_{A_2A_4}^{(01)}={\rm diag}(e^{i\varphi_0},-e^{-i\varphi_0},e^{i(\varphi_2-\varphi_1)},-e^{i(\varphi_1-\varphi_2)}),
  \end{equation}
    \begin{equation}
\hat{U}_{A_2A_4}^{(10)}={\rm diag}(e^{i\varphi_1},-e^{i(\varphi_2-\varphi_0)},-e^{-i\varphi_1},e^{i(\varphi_0-\varphi_2)}),
\end{equation}
    and
\begin{equation}
\hat{U}_{A_2A_4}^{(11)}={\rm diag}(e^{i\varphi_2},e^{i(\varphi_1-\varphi_0)},-e^{i(\varphi_0-\varphi_1)},-e^{-i\varphi_2}).
\end{equation}
Subsequently, Alice measures her qubits $A_2$ and $A_4$ under the a set of
complete orthogonal basis vectors $\{|\pm\rangle:=\frac1{\sqrt2}(|0\rangle\pm|1\rangle)\}$, and broadcasts her measured outcomes
via a classical channel. Incidentally,
all of the authorized participators make an agreement in advance that cbits $(i,j)$ correspond to the outcome
$|{\cal L}_{ij}\rangle_{A_1A_2}$, and cbits $(p,q)$ relate to the measuring outcome of qubits $A_2$ and $A_4$, respectively.
For simplicity, we denote
$$p,q=\left\{
\begin{array}{rcl}
0,      &      & \ {\rm if}\ {|+\rangle}\ {\rm is} \ {\rm probed}\\
1,    &      & \ {\rm if}\ {|-\rangle}\ {\rm is} \ {\rm probed}
\end{array} \right.. $$

${\bf Step\ 3.}$ The agents proceed to carry out single-qubit measurements under the set of vector basis $\{|\pm\rangle\}$ on the qubits respectively, and later
inform Bob of the results via classical channels. We assume that the cbit $x_i$ corresponds to the outcome of the agents $C_i$,
and $y_j$ corresponds to the outcome of the agents $D_j$, where the values of $x_i$ and $y_j$ have been previously denoted as $p$ and $q$, respectively. And we have
$g=\Sigma_{x=1}^nx_i, {\rm mod}\oplus2$ and $h=\Sigma_{y=1}^my_j, {\rm mod}\oplus2$. Actually, there are four different situations,
i.e., I) $g=0$ and $h=0$; II) $g=0$ and $h=1$; III) $g=1$ and $h=0$; and IV) $g=1$ and $h=1$.

${\bf Step\ 4.}$
In response to the different measuring outcomes of the sender and agents, Bob operates on his qubits $B_1$, $B_2$, $B_3$ and $B_4$ with
an appropriate unitary transformation $\hat{U}_{B_1B_2B_3B_4}$.

${\bf Step\ 5.}$ Finally, Bob introduces one auxiliary qubit $B_A$ with initial state of $|0\rangle$. And then he makes
triplet collective unitary transformation $\hat{V}_{B_1B_3B_A}^{(ij)}$ on his qubits $B_1$, $B_3$ and $B_A$ under a set of ordering basis vector $\{|000\rangle,$ $|010\rangle,$ $|100\rangle,$ $|110\rangle,$ $|001\rangle,$ $|011\rangle,$ $|101\rangle,$ $|111\rangle\}$, which is given by
    \begin{equation}
\hat{V}_{B_1B_3B_A}^{(ij)}=\left(
      \begin{array}{cc}
        {\cal W}_{ij}& {\cal U}_{ij}\\
         {\cal U}_{ij} & -{\cal W}_{ij}\\
              \end{array}
    \right),
\end{equation}
where ${\cal W}_{ij}$ and ${\cal U}_{ij}$ are $4\times4$ matrices, respectively. To be explicit, we give
 \begin{equation}
 {\cal W}_{00}={\rm diag}(\frac{a_1b_1}{a_0b_0},\frac{a_1}{a_0},\frac{b_1}{b_0},1),
 \end{equation}
 \begin{equation}
 {\cal U}_{00}={\rm diag}(\sqrt{1-(\frac{a_1b_1}{a_0b_0})^2},
 \sqrt{1-(\frac{a_1}{a_0})^2},\sqrt{1-(\frac{b_1}{b_0})^2},0),
 \end{equation}
 \begin{equation}
 {\cal W}_{01}={\rm diag}(\frac{a_1}{a_0},\frac{a_1b_1}{a_0b_0},1,\frac{b_1}{b_0}),
 \end{equation}
 \begin{equation}
 {\cal U}_{01}={\rm diag}(\sqrt{1-(\frac{a_1}{a_0})^2},
 \sqrt{1-|\frac{a_1b_1}{a_0b_0})^2},0,\! \sqrt{1-(\frac{b_1}{b_0})^2}),
 \end{equation}
 \begin{equation}
 {\cal W}_{10}={\rm diag}(\frac{b_1}{b_0},1,\frac{a_1b_1}{a_0b_0},\frac{a_1}{a_0}),
 \end{equation}
 \begin{equation}
{\cal U}_{10}={\rm diag}(\sqrt{1-|\frac{b_1}{b_0}|^2},0,
 \sqrt{1-(\frac{a_1b_1}{a_0b_0})^2},\sqrt{1-(\frac{a_1}{a_0})^2}),
 \end{equation}
 \begin{equation}
 {\cal W}_{11}={\rm diag}(1,\frac{b_1}{b_0},\frac{a_1}{a_0},\frac{a_1b_1}{a_0b_0}),
 \end{equation}
 and
 \begin{equation}
 {\cal U}_{11}={\rm diag}(0,\sqrt{1-(\frac{b_1}{b_0})^2},\sqrt{1-(\frac{a_1}{a_0})^2},
 \sqrt{1-(\frac{a_1b_1}{a_0b_0})^2}).
 \end{equation}
Then, Bob measures his auxiliary qubit, $B_A$, under a set of measuring basis vectors
$\{|0\rangle,|1\rangle\}$. If state $|1\rangle$ is measured, his remaining
qubits will collapse into the trivial state, and the MJRSP fails in this situation; otherwise,
$|0\rangle$ is probed, and the qubits' state will transform into the desired state, that is, our MCRSP is successful in this case.
\begin{table*}
\caption{$ijpqgh$ denotes the corresponding measurement outcomes from the authorized participants, $\hat{U}_{B_1B_2B_3B_4}$
denotes unitary operations what Bob needs to perform on qubits $B_1$, $B_2$, $B_3$ and $B_4$, respectively.}\footnotesize
\label{tab.1}
\begin{tabular}{|l|l|l|l|l|l|l|l|}
\hline $ijpqgh$ & $\hat{U}_{B_1B_2B_3B_4}$ & $ijpqgh$ & $\hat{U}_{B_1B_2B_3B_4}$ & $ijpqgh$ & $\hat{U}_{B_1B_2B_3B_4}$& $ijpqgh$ & $\hat{U}_{B_1B_2B_3B_4}$\\ \hline
$000000$ & $I_{B1}I_{B2}I_{B3}I_{B4}$ & $010000$ & $I_{B1}I_{B2}X_{B3}X_{B4}$ & $100000$ & $X_{B1}X_{B2}I_{B3}I_{B4}$ &$110000$ & $X_{B1}X_{B2}X_{B3}X_{B4}$ \\
$000001$ & $I_{B1}I_{B2}Z_{B3}I_{B4}$ & $010001$ & $I_{B1}I_{B2}X_{B3}Z_{B3}X_{B4}$& $100001$ & $X_{B1}X_{B2}Z_{B3}I_{B4}$ &$110001$ & $X_{B1}X_{B2}X_{B3}Z_{B3}X_{B4}$ \\
$000010$ & $Z_{B1}I_{B2}I_{B3}I_{B4}$ & $010010$ & $Z_{B1}I_{B2}X_{B3}X_{B4}$ & $100010$ & $X_{B1}Z_{B1}X_{B2}I_{B3}I_{B4}$ &$110010$ & $X_{B1}Z_{B1}X_{B2}X_{B3}X_{B4}$ \\
$000011$ & $Z_{B1}I_{B2}Z_{B3}I_{B4}$ & $010011$ & $Z_{B1}I_{B2}X_{B3}Z_{B3}X_{B4}$ & $100011$ & $X_{B1}Z_{B1}X_{B2}Z_{B3}I_{B4}$ &$110011$ & $X_{B1}Z_{B1}X_{B2}X_{B3}Z_{B3}X_{B4}$ \\
$000100$ & $I_{B1}I_{B2}Z_{B3}I_{B4}$ & $010100$ & $I_{B1}I_{B2}X_{B3}Z_{B3}X_{B4}$ & $100100$ & $X_{B1}X_{B2}Z_{B3}I_{B4}$ &$110100$ & $X_{B1}X_{B2}X_{B3}Z_{B3}X_{B4}$ \\
$000101$ & $I_{B1}I_{B2}I_{B3}I_{B4}$ & $010101$ & $I_{B1}I_{B2}X_{B3}X_{B4}$& $100101$ & $X_{B1}X_{B2}I_{B3}I_{B4}$ &$110101$ & $X_{B1}X_{B2}X_{B3}X_{B4}$ \\
$000110$ & $Z_{B1}I_{B2}Z_{B3}I_{B4}$ & $010110$ & $Z_{B1}I_{B2}X_{B3}Z_{B3}X_{B4}$ & $100110$ & $X_{B1}Z_{B1}X_{B2}Z_{B3}I_{B4}$ &$110110$ & $X_{B1}Z_{B1}X_{B2}X_{B3}Z_{B3}X_{B4}$ \\
$000111$ & $I_{B1}I_{B2}Z_{B3}I_{B4}$ & $010111$ & $Z_{B1}I_{B2}X_{B3}X_{B4}$ & $100111$ & $X_{B1}Z_{B1}X_{B2}I_{B3}I_{B4}$ &$110111$ & $X_{B1}Z_{B1}X_{B2}X_{B3}X_{B4}$ \\
$001000$ & $Z_{B1}I_{B2}I_{B3}I_{B4}$ & $011000$ & $I_{B1}I_{B2}X_{B3}Z_{B3}X_{B4}$ & $101000$ & $X_{B1}Z_{B1}X_{B2}I_{B3}I_{B4}$ &$111000$ & $X_{B1}Z_{B1}X_{B2}X_{B3}X_{B4}$ \\
$001001$ & $Z_{B1}I_{B2}Z_{B3}I_{B4}$ & $011001$ & $Z_{B1}I_{B2}X_{B3}Z_{B3}X_{B4}$& $101001$ & $X_{B1}Z_{B1}X_{B2}Z_{B3}I_{B4}$ &$111001$ & $X_{B1}Z_{B1}X_{B2}X_{B3}Z_{B3}X_{B4}$ \\
$001010$ & $I_{B1}I_{B2}Z_{B3}I_{B4}$ & $011010$ & $I_{B1}I_{B2}X_{B3}X_{B4}$ & $101010$ & $X_{B1}X_{B2}I_{B3}I_{B4}$ &$111010$ & $X_{B1}X_{B2}X_{B3}X_{B4}$ \\
$001011$ & $I_{B1}I_{B2}I_{B3}I_{B4}$ & $011011$ & $I_{B1}I_{B2}X_{B3}Z_{B3}X_{B4}$ & $101011$ & $X_{B1}X_{B2}Z_{B3}I_{B4}$ &$111011$ & $X_{B1}X_{B2}X_{B3}Z_{B3}X_{B4}$ \\
$001100$ & $Z_{B1}I_{B2}Z_{B3}I_{B4}$ & $011100$ & $Z_{B1}I_{B2}X_{B3}Z_{B3}X_{B4}$ & $101100$ & $X_{B1}Z_{B1}X_{B2}Z_{B3}I_{B4}$ &$111100$ & $X_{B1}Z_{B1}X_{B2}X_{B3}Z_{B3}X_{B4}$ \\
$001101$ & $Z_{B1}I_{B2}I_{B3}I_{B4}$ & $011101$ & $Z_{B1}I_{B2}X_{B3}X_{B4}$& $101101$ & $X_{B1}Z_{B1}X_{B2}I_{B3}I_{B4}$ &$111101$ & $X_{B1}Z_{B1}X_{B2}X_{B3}X_{B4}$ \\
$001110$ & $I_{B1}I_{B2}I_{B4}$ & $011110$ & $I_{B1}I_{B2}X_{B3}Z_{B3}X_{B4}$ & $101110$ & $X_{B1}X_{B2}Z_{B3}I_{B4}$ &$111110$ & $X_{B1}X_{B2}X_{B3}Z_{B3}X_{B4}$ \\
$001111$ & $I_{B1}I_{B2}I_{B3}I_{B4}$ & $011111$ & $I_{B1}I_{B2}X_{B3}X_{B4}$ & $101111$ & $X_{B1}X_{B2}I_{B3}I_{B4}$ &$111111$ & $X_{B1}X_{B2}X_{B3}X_{B4}$ \\
\hline
\end{tabular}
\end{table*}

Based on the above five-step protocol, it has been shown that the MCRSP for a family of cluster-type states
can be faithfully performed with predictable probability. The steps can be decomposed into a
schematic diagram shown in Fig.~\ref{fig.1}.
 As a summary, we list Bob's required local single-qubit transformations according to the sender's
and agents' different measurement outcomes in Table~\ref{tab.1}.

From the above analysis, one can see that the prepared state can be faithfully reconstructed with
specified success probabilities.

Now, let us turn to calculate the TSP and CCC.
Alice's measurement outcome, $|{\mathcal{L}}_{ij}\rangle$, has an occurrence probability of
\begin{equation}
P_{|{\mathcal{L}}_{ij}\rangle}=\frac1{{\cal N}_{ij}^2}.
 \end{equation}
Furthermore, in considering the capture of the state $|0\rangle_{B_A}$, the probability should be
\begin{equation}
 P_{|0\rangle_{B_A}}={|{\cal N}_{ij}a_1b_1|}^2.
 \end{equation}
Thus, the success probability of MCRSP for the measurement outcome $(i,j)$ should be given by
\begin{equation}
 P_{(i,j)}=P_{|{\mathcal{L}}_{ij}\rangle}\times{P}_{|0\rangle_{B_A}}=|a_1b_1|^2.
 \end{equation}
In terms of $P_{(i,j)}$, one can easily obtain that the TSP sums to
 \begin{equation}
 P_{\sum_{i,j}^{0,1}(i,j)}=\sum_{i,j}^{0,1}{P}_{(i,j)}=4|a_1b_1|^2.
 \end{equation}

Moreover, one can show that the required CCC
should be $(2+2+m+n)=(m+n+4)$ cbits totally.

\begin{figure}
\includegraphics[width=11cm,height=4.6cm]{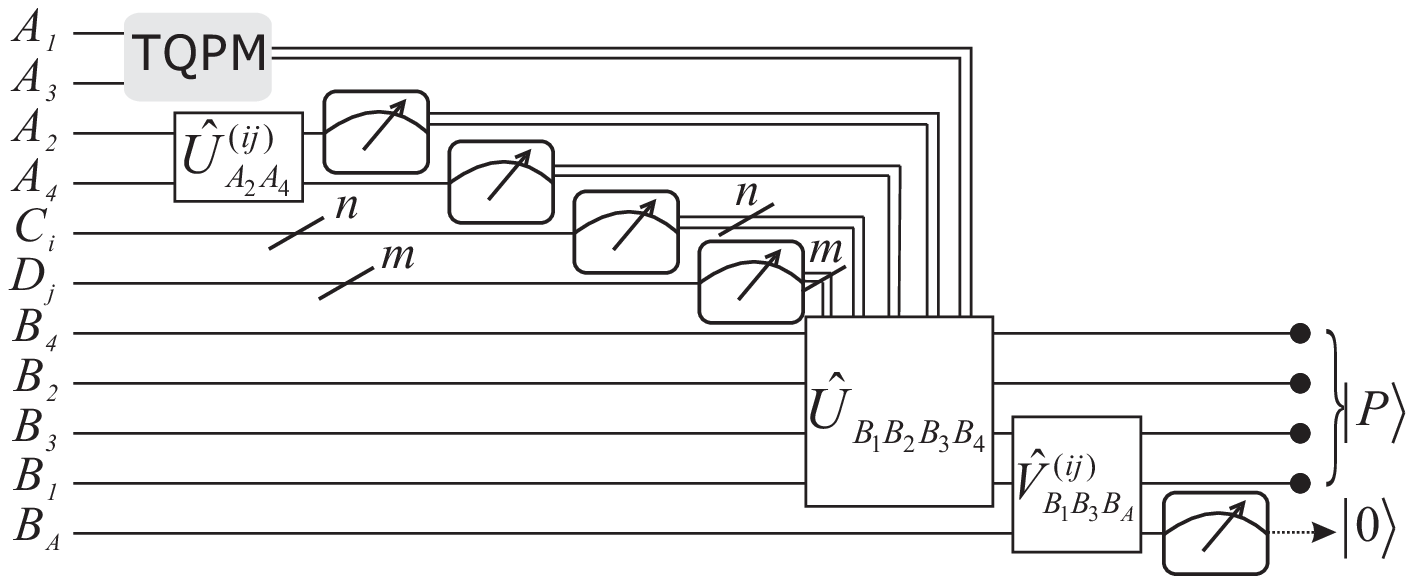}
\caption{Quantum circuit for implementing the MCRSP scheme. TQPM denotes two-qubit projective measurement under
a set of complete orthogonal basis vectors $\{|{\cal L}_{ij}\rangle\}$; $\hat{U}_{A_2A_4}^{(ij)}$ denotes Alice's
appropriate bipartite collective unitary transformation on qubit pair ($A_2$, $A_4$); $\hat{U}_{B_1B_2B_3B_4}$ denotes Bob's
appropriate single-qubit unitary transformations on his qubits $B_1$, $B_2$, $B_3$ and $B_4$, respectively, and
$\hat{V}_{A_1A_3B_A}^{(ij)}$ denotes Bob's
triplet collective unitary transformation on his qubits $B_1$, $B_3$ and $B_A$.}
\label{fig.2}
\end{figure}

Herein, we had described our proposal of MCRSP for a family of four-qubit cluster-type entangled states.
We have proved that our scheme can  be realized faithfully with TSP of $4|a_1b_1|^2$ and CCC of
$(m+n+4)$ via the control of multi-agent in a quantum network. For clarity, the quantum circuit for our
MCRSP protocol is displayed in Fig.~\ref{fig.2}.
\begin{figure}
\centering
\includegraphics[height=6cm]{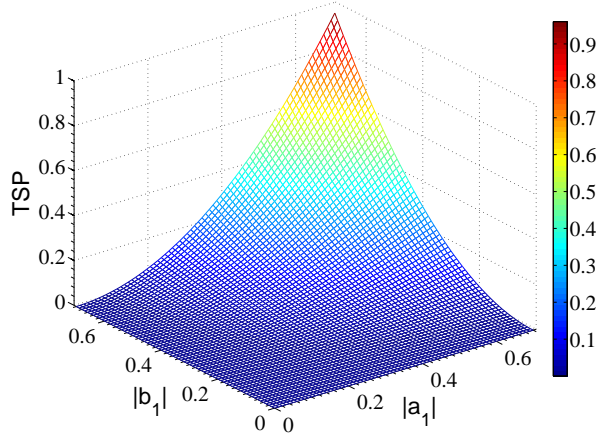}
\caption{The relation between TSP and the smaller coefficients of entanglements severed as quantum channels.}
\label{fig.3}
\end{figure}
\section{Discussions}
\begin{figure}
\centering
\includegraphics[width=7cm,height=5.5cm]{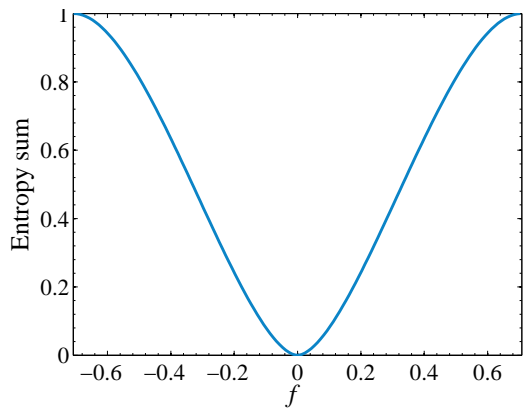}
\caption{The entropic diagram with variation of the parameter of quantum channels.}
\label{fig.4}
\end{figure}

We have found several remarkable features with respect to the scheme presented above and these features are summarized as follows:
 (1) To the  best of our knowledge, this is the first time one
has exploited such a scenario concerning MCRSP for four-qubit cluster-type entangled states
via control of $(m+n)-$party. Information
conveyance only takes place between the sender and the receiver, i.e., $1\rightarrow1$ threshold communication.
Moreover, the agents are capable of
supervising and switching the procedure during the relay of information communication. Secure multi-node information
communication is considerably important in prospective quantum networks.
(2) Generally, our MCRSP can be faithfully performed with TSP of $4|a_1b_1|^2$.
 Moreover, when the state $|a_1|=|b_1|=1/\sqrt2$ is chosen in the beginning, thus the channels become maximally entangled,
 the TSP can reach unity as shown in Fig.~\ref{fig.3}. Consequently, that indicates our scheme becomes a deterministic one in this case.
Additionally, it should be noted that the parameters $a_1$ and $b_1$ relate to the Shannon entropies of the employed quantum channels,
 \begin{equation}
 H(f)=-|f|^2{\rm log}|{f}|^2-(1-|f|^2){\rm log}(1-|{f}|^2),
 \end{equation}
 where $f\in\{a_1,b_1\}$ and $a_1,b_1\in[-\sqrt2/2,\sqrt2/2]$. The entropy will vary with the coefficients specific to
different quantum channels depicted in Fig.~\ref{fig.4}. Note, the entropy in essence reflects
inherent property (i.e., entanglements) of quantum channels.
(3) Our scheme enables one to fulfill RSP via the multi-agent control. Incidentally, all of the agents are capable of switching
the preparation procedures. The desired state can be recovered at Bob's site conditioned to the total collaboration of network members.
Anyone of the party cannot recover the desired state by themselves. In this sense,
the security of information is to a large extend guaranteed.
(4) Within our scheme, there exists $(m+n)$ controllers to manipulate or switch the preparation procedure.
If both $m$ and $n$ are chosen to be 0, there are no authorized controllers during the process of the preparation,
it has been found that our scheme is smoothly reduced to a scheme resembling RSP for four-qubit cluster-type states with TSP of $4|a_1b_1|^2$.
In this case, the measurements made by the controllers and the communication between controllers and receivers are unnecessary,
as is the auxiliary qubit. Now, we can compare our reduced
scheme with other previous schemes~\cite{Y.B,D.Wan6,D.Wang5,K.Hou333,Y.B.11,K.Hou}; we do this
with respect to RSP and JRSP for such states in view of the resource consumption and quantum operation complexity as shown in Table~\ref{tab.2}.\\

\begin{table*}
\caption{Comparison between our scheme and the previous ones in the case of
maximally entangled channels.
ET represents entanglement; SQ represents single-qubit; ASQ represents auxiliary single-qubit; CNOT represents controlled-not gates;
PM represents projective measurement;
SQPM represents single-qubit projective measurement and TSP represents total success probability.}
\begin{center}
\label{tab.2}
\begin{tabular}{|c|l|l|c|c|c|}
\hline Schemes & Required qubits & Quantum operations & CCC & TSP & $\eta$\\ \hline
Ref. \cite{Y.B} & six 2-qubit ETs & two 4-qubit PMs & 8 & $\frac1{16}$& 1.25\% \\
 & two 6-qubit ETs &  two 4-qubit PMs & 8 & $\frac1{16}$ &  1.25\% \\
Ref. \cite{D.Wan6} & two 4-qubit ETs & two 2-qubit PMs & 4 &  $\frac14$ & 8.33\%  \\
Ref. \cite{D.Wang5} & two 3-qubit ETs \& two ASQ & two 2-qubit PMs \& 2 CNOTs & 4 &  $\frac14$ & 8.33\% \\
Ref. \cite{K.Hou333} & two 2-qubit ETs \& four ASQ & two 4-qubit PMs \&  4 CNOTs & 4 &  $\frac14$ & 8.33\%  \\
Ref. \cite{Y.B.11} & six 2-qubit ETs & two 4-qubit PMs & 8 &  1 & 20.00\% \\
Ref. \cite{K.Hou} & two 2-qubit \& one 3-qubit ET & one 3-qubit PM & 3 &  $\frac14$ & 10.00\% \\
Current scheme & two 4-qubit ETs & one 2-qbuit PMs \& two SQPMs &4&  1 & 33.33\% \\
\hline
\end{tabular}
\end{center}
\end{table*}
From Table~\ref{tab.2}, one can directly note that the TSP of our scheme is capable of unity, and the intrinsic efficiency $(\eta)$
achieves $33.33\%$, which is much greater than those in the previous schemes \cite{Y.B,D.Wan6,D.Wang5,K.Hou333,Y.B.11,K.Hou}.
Due to characteristic high-efficiency and high-TSP in the present scheme, it is both highly efficient and optimal in comparison to the
existed ones.
Incidentally, the intrinsic efficiency of a scheme is defined by \cite{Yuan}
 \begin{equation}
 \eta=\frac{N_s}{N_q+N_c}\times{TSP},
 \end{equation}
where $N_s$ weights the amount of qubits of the prepared states, $N_q$ weights the amount of quantum resource consumption,
and $N_c$ weights the amount of CCC in quantum computation. Additionally, Ref. \cite{Y.B.11} can be realized with a TSP of $100\%$;
however, there are several crucial differences between our methods and the previous, they are as follows: (i) Quantum resource consumption. In \cite{Y.B.11}, 12 qubits are indispensable in the course of RSP for four-qubit cluster-type states, while 8 qubits are sufficient to implement RSP for such states in our reduced scheme.
Implying our scheme is more economic. (ii) Operation complexity. Two four-qubit projective measurements in \cite{Y.B.11} are require for their procedure, while two-qubit projection measurements are required in our scheme. Experimental realization of four-qubit projective measurement is much more difficult than that for two-qubit. Thus, in principal our scheme is easier to experimentally realize than the previous method.

\section{Conclusion}
Herein we have derived a novel strategy for implementing MCRSP scheme for a family of
four-qubit cluster-type entangled states by taking advantage of robust GHZ-class states as quantum channels.
With the aid of suitable LOCC, our scheme can be realized with high success probability. Remarkably,
our scheme has several nontrivial features, including high success probability, security and reducibility.
Particularly, the TSP of MCRSP can reach unity when the quantum channels are distilled to maximally entangled ones;
that is, our scheme can be performed deterministically at this limit.
We argue the current MCRSP proposal might open up a new way for long-distance communication
in prospective multi-node quantum networks.

\acknowledgments
This work was supported by NSFC (11247256, 11074002, and 61275119), the fund of Anhui Provincial Natural Science Foundation,
the fund of China Scholarship Council and project from National Science Foundation Centers for Chemical Innovation:
CHE-1037992.

\end{document}